# Effect of Na doping on photoluminescence and laser stimulated nonlinear optical features of ZnO nanostructures

U.G. Deekshitha[a], Krithika Upadhya[a], Albin Antony[a], Aninamol Ani[a], M. Nowak[b], I.V. Kityk[b], J. Jedryka[b], P. Poornesh[a],*, K.B. Manjunatha[c],**, Suresh D. kulkarni[d] a *Department of Physics, Manipal Institute of Technology, Manipal Academy of Higher Education, Manipal, Karnataka, 576104, India*
b *Institute of Optoelectronics and Measuring Systems, Faculty of Electrical Engineering, Czestochowa University of Technology, PL-42-201, ArmiiKrajowej 17, Czestochowa, Poland*
c *Department of Physics, NMAM Institute of Technology, Nitte, 574110, India*
d *Department of Atomic and Molecular Physics, Manipal Academy of Higher Education, Manipal, Karnataka, 576104, India*

ABSTRACT

In this work, nonlinear optical properties of Na: ZnO thin films (Na: ZnO) have been experimentally elaborated. The principal possibility to operate the nonlinear optical features using external laser beams is shown. The Na: ZnO films were synthesized by spray pyrolysis technique at a deposition temperature equal to about 400 °C. XRD graph reveals that the grown films were polycrystalline in nature with a dominant peak corresponding to (0 0 2) plane. Despite the difference in the ionic radii of the Na ($0.95 A°$) and Zn ($0.74 A°$), an angle shift in the XRD peak was not observed, whereas there was a significant change in peak intensity. The photoluminescence (PL) spectra resulted in three emission centres spectrally situated in violet, blue and green colour region due to the presence of native defect states in the forbidden energy gap. Second and third harmonic generation (SHG, THG) ex-periments stimulated by external coherent light beams show the existence of THG maxima for 15% Na doped Zn:O which was quite different with respect to SHG maximum for pure ZnO. The main innovation of this work is the possibility to change nonlinear optical susceptibility varying Na doping concentration and by coherent laser treatment, contrary to the previous works where these parameters have not been explored.

## 1. Introduction

Recently there has been observed an enhanced interest on Zinc Oxide (ZnO) based nanostructure because of their large variety of applications mainly in field-effect transistors, piezoelectric nano-generators, photodiode, solar cell, coherent laser transformation, gas sensors etc…[1]. These materials may be applied as photocatalyst and sensors due to their promising piezoelectric and optical and nonlinear optical features [2–6]. Large exciton binding energy of 60 meV, a wide direct band gap about 3.37eV, nontoxic, and a relatively high melting point of 1975 °C, easy device fabrication, high optical transmittance, high thermal conductivity and high electron mobility has always made ZnO as a promising candidate for various applications [1–9]. In order to enhance and modulate the structural and optical properties of ZnO thin film to meet the ever-increasing technological demand, many approaches have been implemented. Among all, doping with a suitable element which can bring out substantial variation in physical and chemical properties are one of the widely used methods [10–14]. In the present study, we incorporate alkali metal Sodium (Na) with atomic weight 22.99u, density 0.971 g/cm as the doping material. Na belongs to Group I and is a suitable acceptor with relatively high hole concentration [2,8].

Many techniques are reported for the synthesis of Na: ZnO thin films including DC and RF sputtering, thermal evaporation, chemical vapour deposition, spin coating, spray pyrolysis, sol-gel etc. Among this spray, pyrolysis has gained a lot of attention because of its low cost, large area deposition, safe and simple deposition methodology [8]. In this work, pure and Na doped ZnO thin films have been prepared by spray pyr-olysis technique at different Na doping concentration. We have in-corporated Z-scan technique using a continuous laser which accounts the thermal nonlinearity of the sample and THG, SHG technique using a nanosecond laser which accounts the electronic nonlinearity of the sample. The studies which combined both of these domains are not well explored till date. Additionally, in this work, we present the principal possibility to operate the nonlinear optical features using external laser beams which is contrary to most of the previous work where the nonlinear optical parameters are fixed. The main innovation of this work is to explore the possibility to change nonlinear optical susceptibilities by Na doping concentration and coherent laser treatment differing to the previous works where these parameters have not been explored.

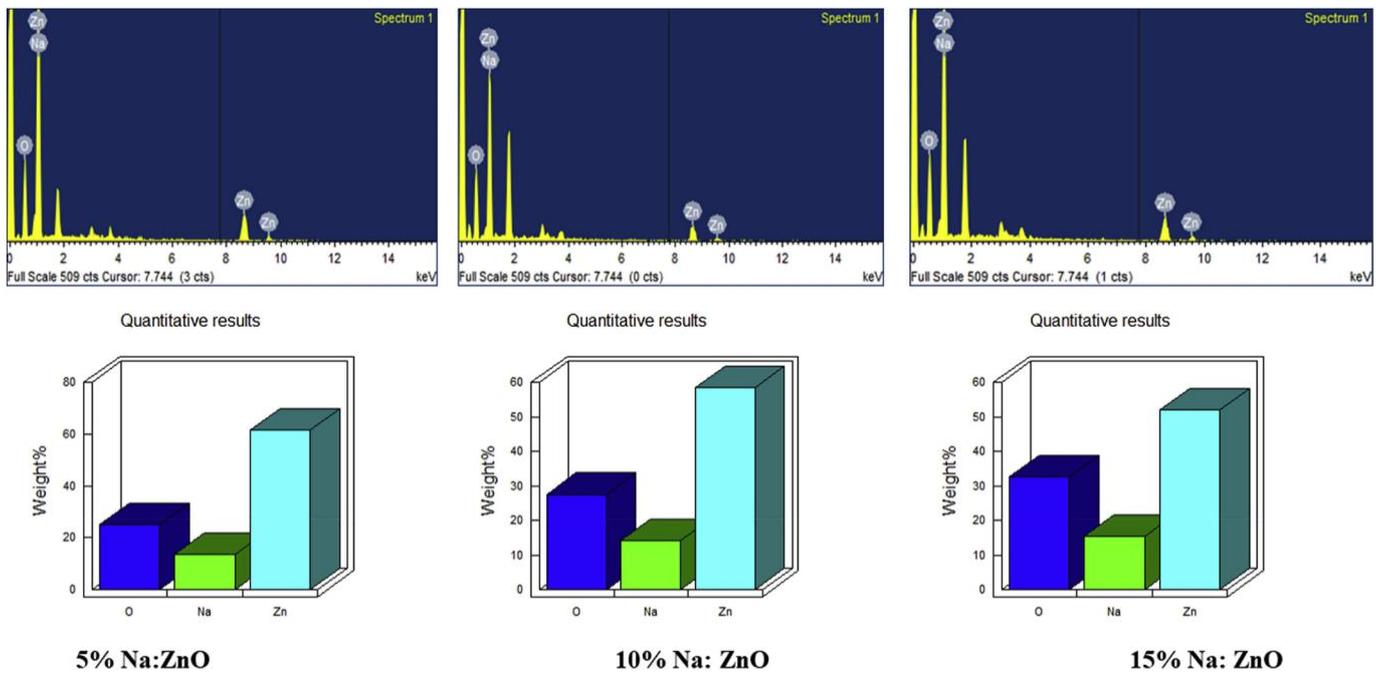

**Fig. 1.** EDS analysis of Na: ZnO thin films.

## 2. Experimental details

### 2.1. Thin film preparation by spray pyrolysis

The Na: ZnO thin films have been fabricated using a low-cost spray pyrolysis technique. Sodium acetate [$CH_3COONa \cdot 3H_2O$, $M_w$ = 136.08] and Zinc acetate [$Zn(CH_3COO)_2 \cdot 2H_2O$, $M_w$ = 219.49] has been served as a precursor for sodium (dopant) and zinc (host) to prepare pure and Na doped ZnO thin films. Sodium doping was done by addition of sodium acetate at a doping concentration of 5%, 10% and 15% to Zn precursor. The working temperature was maintained at 400 °C. The elemental composition of the grown films was then confirmed by Energy-dispersive X-ray spectroscopy analysis. Fig. 1 shows the EDS spectra and quantitative analysis of the grown films at different Na doping concentration.

### 2.2. Characterization techniques

The surface morphology and structural properties of grown films were investigated by atom force microscopy (AFM) and X-ray diffraction (XRD). JASCO Fluorescence spectrometer with a spectral resolution of 1 nm was employed to record the photoluminescence spectroscopy of the grown films. Xenon flash lamp was used as the probing source with an excitation wavelength of 325 nm to explore the radiative defects in the films. Open aperture Z-scan measurements were used to determine the sign and magnitude of third-order nonlinear absorption coefficient ($\beta_{eff}$) for Na doped ZnO thin films. He–Ne laser of 632 nm wavelength and 22 mW input power were used as the excitation source. The samples were placed on an automated translational stage and move along the focus of the excitation source at a distance of 40 mm. The variations in the output power at an increment of 500 μm were recorded by a photodetector and further analysed using nonlinear transmission equations. Laser-induced second and third harmonic generation techniques with a fundamental source of 10 ns Nd: YAG laser emitting at 1064 nm with pulse duration 10 ns and pulse frequency repetition 10 Hz were employed for second order and third order nonlinear optical property analysis. Linear optical properties of the films were studied using SHIMADZU UV 1800 spectrophotometer within the wavelength range of 350–1000 nm.

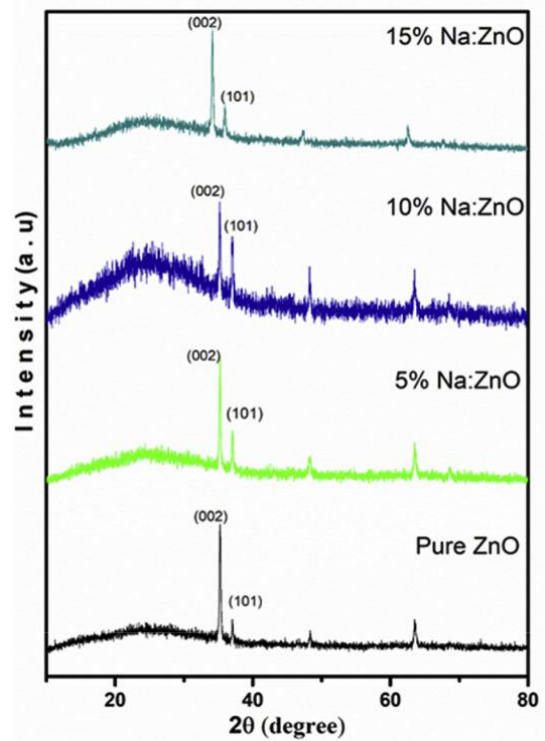

**Fig. 2.** XRD spectra of pure and Na: ZnO thin films.

| Table 1 |
|---|
| Structural parameters of the Na: ZnO films. |

| Dopant Na (Wt. %) | Crystallite size (D) (nm) | Dislocation density (δ) $10^{14}$ lines/m$^2$ | Strain(ε) $10^{-3}$ |
|---|---|---|---|
| 0 | 34.98 | 8.16 | 0.99 |
| 5 | 36.54 | 7.48 | 0.94 |
| 10 | 29.75 | 11.29 | 1.16 |
| 15 | 37.76 | 7.01 | 0.91 |

| Table 2 |
|---|
| Surface Roughness of the Na: ZnO films. |

| Dopant Na (Wt. %) | Surface Roughness (nm) |
|---|---|
| 0% | 16.8 |
| 5% | 23.2 |
| 10% | 24.2 |
| 15% | 33.8 |

## 3. Results and discussion

### 3.1. Structural and morphological studies

X-Ray diffraction (XRD) patterns of pure and Na: ZnO thin films are depicted in Fig. 2. XRD graph shows multiple peaks confirming a fact that the titled films are confirming successful doping of Na into ZnO lattice [1]. The dominant peak was observed along (0 0 2) plane [1]. Doping of Na doesn't result in the shifting of XRD peak despite differences in ionic radii of Na (0.95A°) and Zn (0.74A°) [2] but amended XRD peak intensity. Furthermore, large differences in the ionic radii resulted in the incorporation of lattice defects which is evident from the variation in dislocation density.

The average crystallite sizes were calculated from (0 0 2) peak using Debye's Scherer formula [3]:

$$D = \frac{K\lambda}{\beta \cos\theta} \quad (1)$$

where $\theta$, $\lambda$, $k$, $D$ are the Bragg diffraction angle, the X-ray wavelength ($\lambda$ = 1.5406 A°), FWHM, shaping factor, average crystal size, respectively [1]. Value of the dislocation density was determined using Eq. 2

$$\delta = \frac{1}{D^2} \quad (2)$$

Using Eq. (3) strain-induced effects in the lattice upon Na doping were calculated.

$$\epsilon = \frac{\beta \cos\theta}{4} \quad (3)$$

When the dopant Na is introduced into ZnO lattice, the crystalline sizes have been varied. This indicates that crystallinity of the films modifies as the doping concentration increases and the lower value of dislocation density and strain indicates the good crystalline quality of the grown films [9]. The calculated structural parameters of Na: ZnO nanostructures are tabulated in Table 1.

Fig. 3 presents the AFM image of the pure and Na: ZnO films. The surface roughness values are shown in Table 2. As the doping concentration increases, surface roughness also shows an enhancement. It is noteworthy that at the loose surface, voids exist and as the Na concentration increases there will be the formation of larger grains

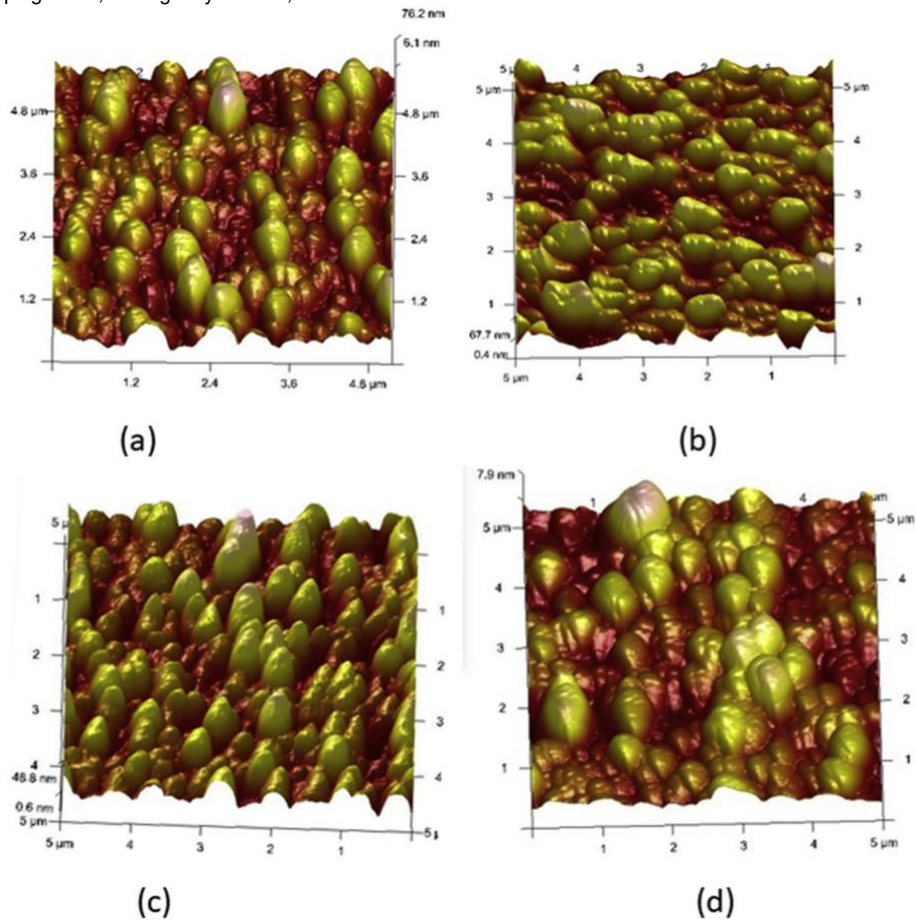

**Fig. 3.** AFM images of (a) pure ZnO (b) 5% (c) 10% (d) 15% Na:ZnO films.

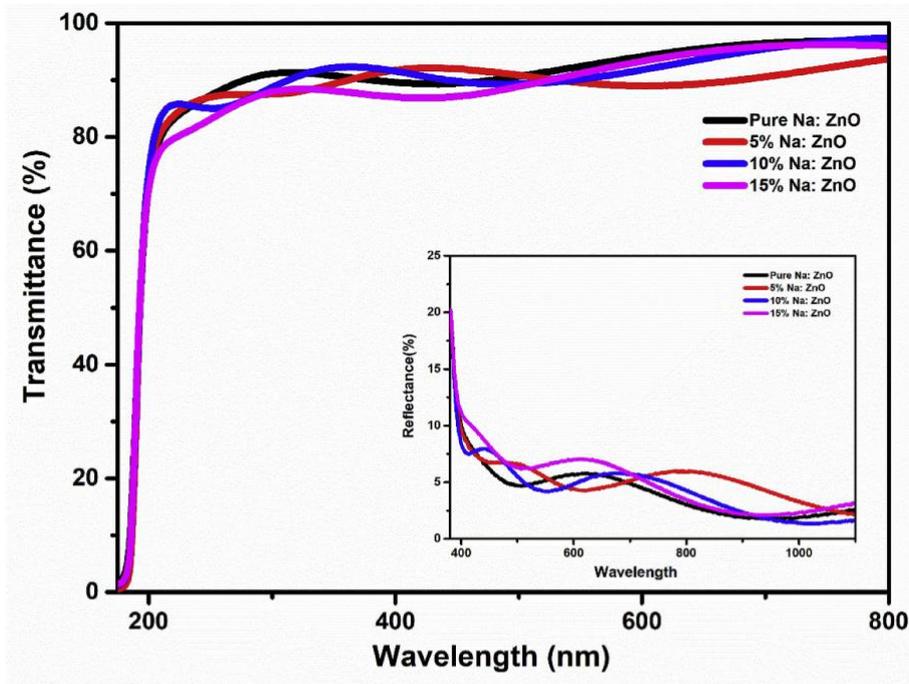

a

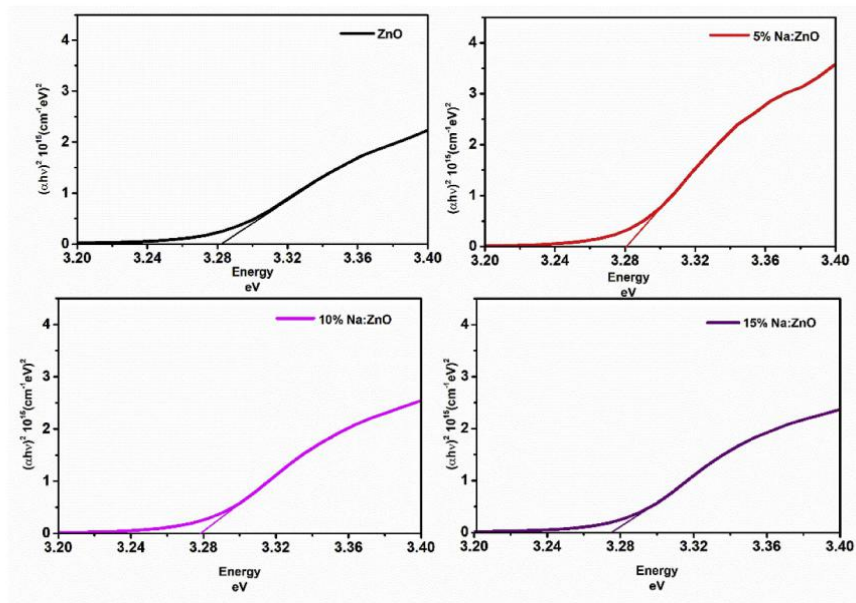

b

**Fig. 4.** a Transmittance and Reflectance (inset) spectra of Na: ZnO thin films. (b) Tauc Plot analysis of Na: ZnO films.

resulting in an enhancement of average surface roughness [8].

*3.2. Optical studies*

UV Visible spectrophotometer with spectral resolution 1 nm was used to study the optical properties of Na: ZnO films within the spectral range 300–1000 nm. The optical transmittance and reflectance graph (inset) of Na: ZnO thin films are shown inFig. 4. (a). From the figure it is observed that optical reflectance from the film surface compratively less and major part of the incident light is transmiited. This observed behaviour indicates that the grown films depicts negligible fabry-perot effect.

By extrapolating the linear part of the Tauc plot shown in Fig 4b will give the values of the energy band gap (Eg) of the film [9]. The band gap for pure and Na: ZnO thin films was almost same and shows a slight variation. The observed energy shift in the band gap was attributed to the alteration of the lattice constant of ZnO nanostructures. Moreover, after Na doping defects are generated in the nanostructures [2] and for the higher concentration of Na: ZnO, impurity levels near the conduc-tion band has a minimum and valence band a maximum [8]. Thus, additional band tail arose resulting in an observed reduction of the energy band gap [7].

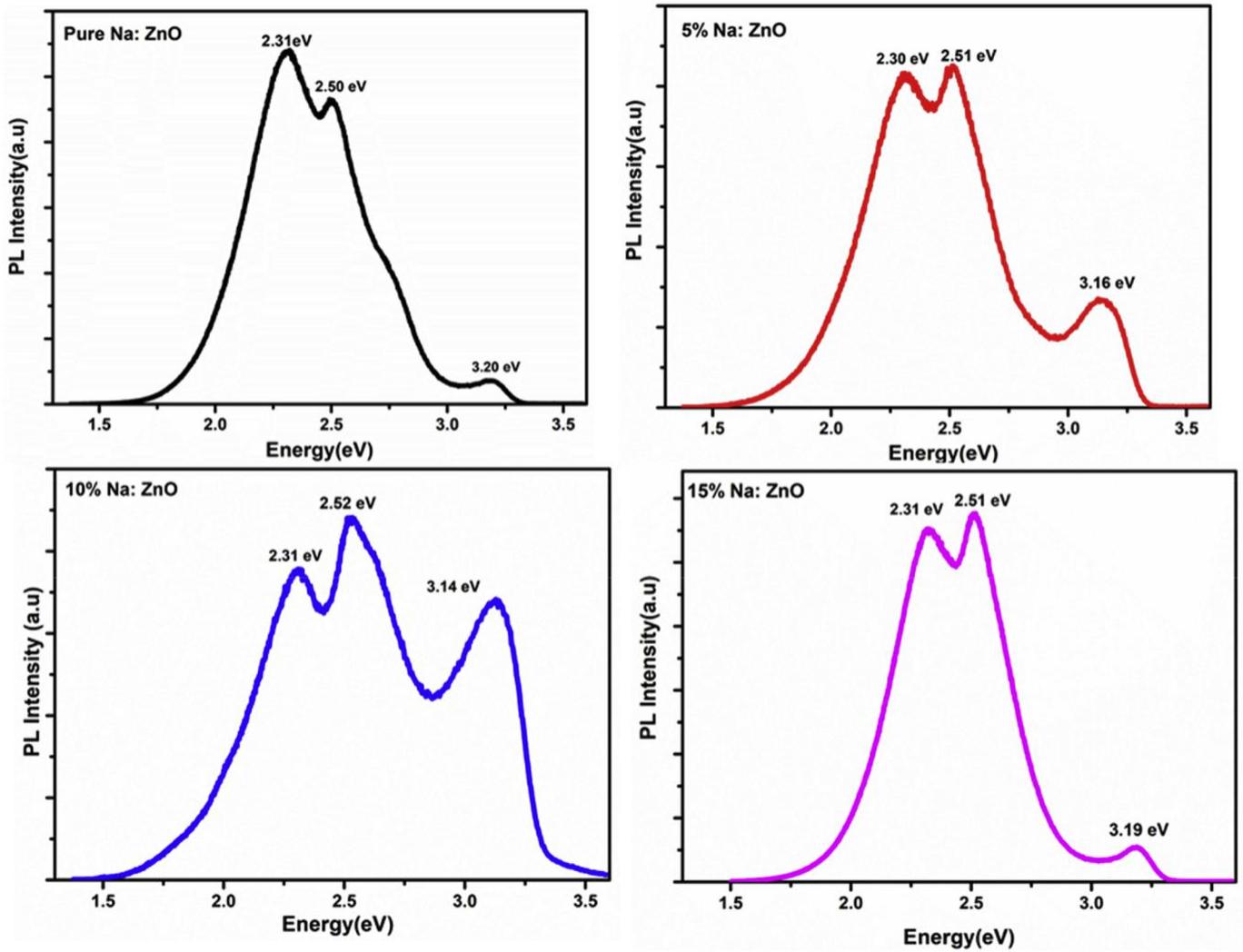

**Fig. 5.** PL spectra of Na: ZnO at different Na concentration.

### 3.3. Photoluminescence properties

The ambient temperature PL spectra of pure and Na: ZnO thin films were obtained at an excitation wavelength of 325 nm. Fig. 5 shows the PL spectra of Na: ZnO thin film for pure, 5%, 10%, 15% concentration. The observed broad spectra appeared due to the presence of re-combination sites, grain boundaries and defects [15]. Generally, ZnO thin film is presented with the native defect such as zinc vacancies, zinc interstitial, oxygen vacancy and absorbed hydroxyl group [15]. The near band edge emission observed at energies around 3.20 eV, 3.16 eV, 3.12 eV, 3.19 which were due to the transition of an electron from the conduction band to an ionised oxygen vacancy in the valence band [16–19]. The blue emission observed around ~2.51eV for Na: ZnO films were caused by the oxygen antisites defect sites (Ozn) which re-sults in the potential electron transition from the conduction band to Ozn defect site [16–19]. In Zn rich environment, the formation of Ozn defect is very typical and the intensity of blue emission varied upon doping. The green emission observed at 2.31eV is attributed to the recombination of holes and electrons in the deep level oxygen vacancy (Vo) and electron trapped in a shallow level situated below conduction band of Na: ZnO thin film [18,19]. Table 3 summarizes the observed and PL emission centres and its proposed origin in Na: ZnO thin films.

### 3.4. Nonlinear optical properties

#### 3.4.1. Open aperture Z-scan measurements

Third-order nonlinear absorption coefficients of the films were measured by the Z-scan technique as described by Sheikh Bahae et al. [20]. He–Ne laser of 632 nm wavelength, 20 mW input power and $TEM_{00}$ mode was applied for the third order nonlinear optical measurement described by fourth rank tensor. Applying the open aperture Z-scan technique we have evaluated the nonlinear absorption coefficient $\beta_{eff}$ of the films. Fig. 6 shows the open aperture Z-scan trace for the studied Na: ZnO thin films. From the trace, it is observed that pure, 10% and 15% Na: ZnO thin films have positive absorption non-linearity, which corresponds to reverse saturable absorption mechanism (R.S.A) [21]. For 5% Na: ZnO thin film saturable absorption mechanism (S.A) was perceived which corresponds to negative absorption non-linearity. The switching over mechanism from RSA to SA is due to the fact that absorption in the excited state is higher with respect to the ground state and further non-availability of empty defect states [19,22].

**Table 3**
PL emission centres and their proposed origin.

| Pure | | 5% Na: ZnO | | 10% Na: ZnO | | 15% Na: ZnO | |
| --- | --- | --- | --- | --- | --- | --- | --- |
| Energy(eV) | Defects | Energy(eV) | Defects | Energy(eV) | Defects | Energy(eV) | Defects |
| 3.20 | NBE | 3.16 | NBE | 3.14 | NBE | 3.19 | NBE |
| 2.50 | Ozn | 2.51 | Ozn | 2.52 | Ozn | 2.51 | Ozn |
| 2.31 | Vo | 2.30 | Ozn | 2.31 | Vo | 2.31 | Vo |

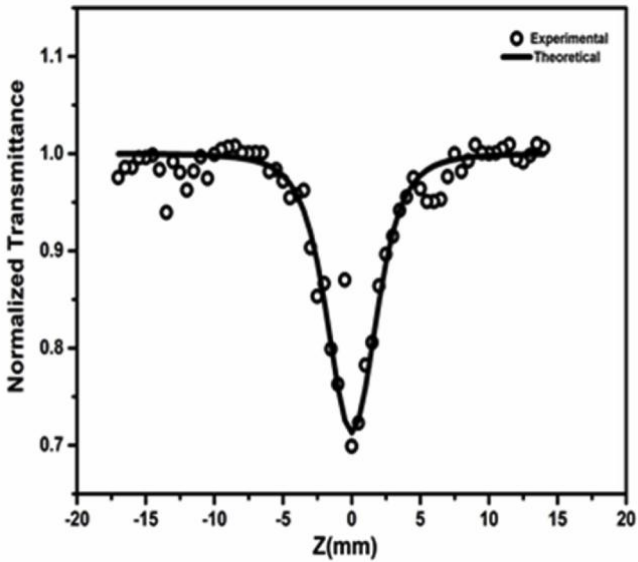 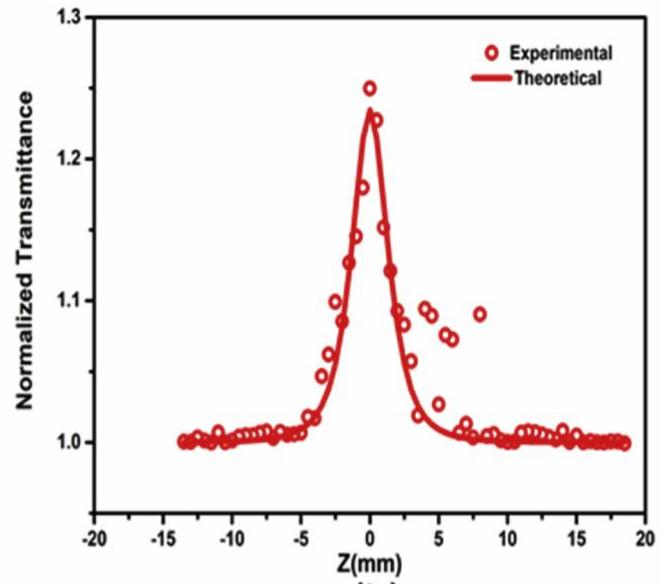

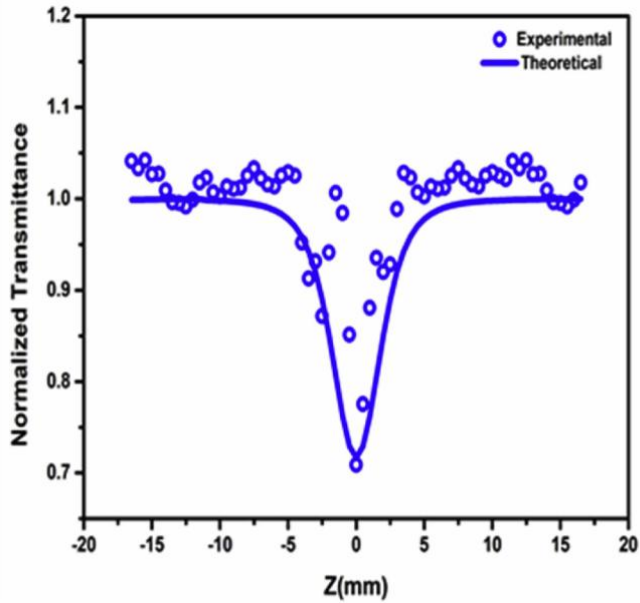 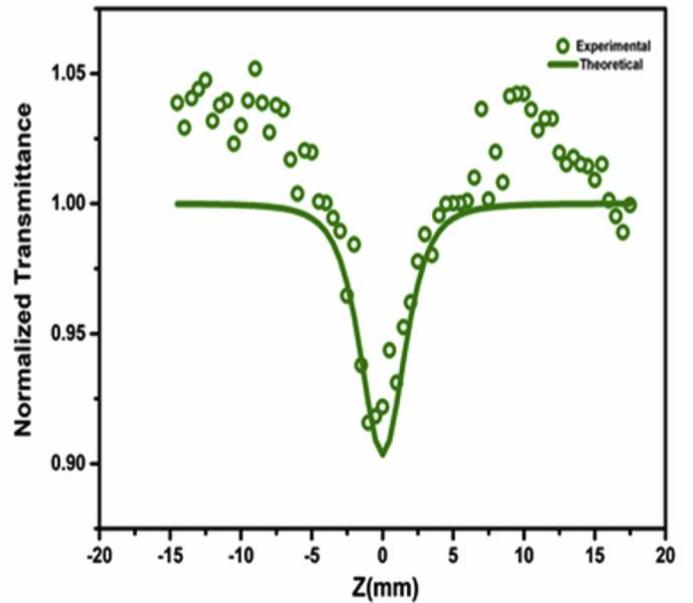

**Fig. 6.** Open aperture Z-scan trace of Na: ZnO thin films.

**Table 4**
Shows the absorption coefficient of Na: ZnO for dif-ferent concentration.

| Conc. (%) | $B_{eff}$ ($10^{-1}$) cm/W |
|---|---|
| ZnO | 2.90 |
| 5% ZnO | −2.27 |
| 10%ZnO | 2.84 |
| 15% ZnO | 0.71 |

**Table 5**
Reported third-order nonlinear absorption coefficient of Semicondutor mate-rials [23–27].

| Material | $\beta_{eff}$(Cm/W) |
|---|---|
| Fluorine doped ZnO | $-0.371 \times 10^{-2}$ |
| Sr doped ZnO PVA thin films | $-6.57 \times 10^{-4}$ |
| ZnS nanoparticle | $-3.2 \times 10^{-3}$ |
| Mn doped ZnS nano particles | $-4.7 \times 10^{-3}$ |
| ZnO: Sn | $-0.97 \times 10^{-3}$ - $-9.6 \times 10^{-3}$ |

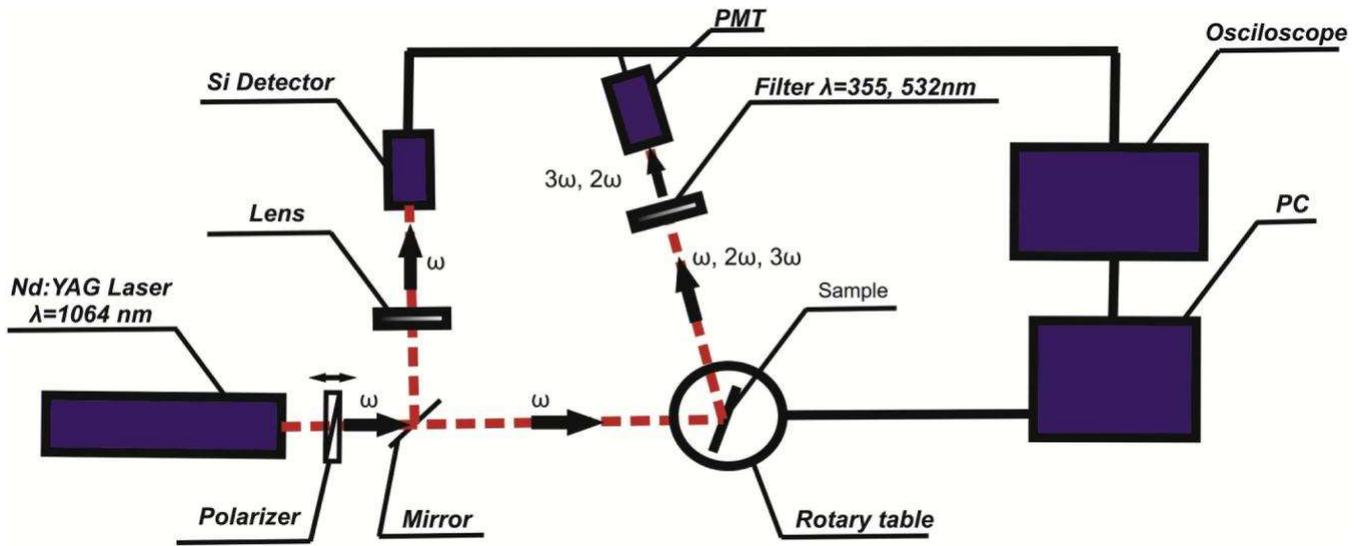

**Fig. 7.** Principal set-up for the measurements of the laser stimulated SHG and THG.

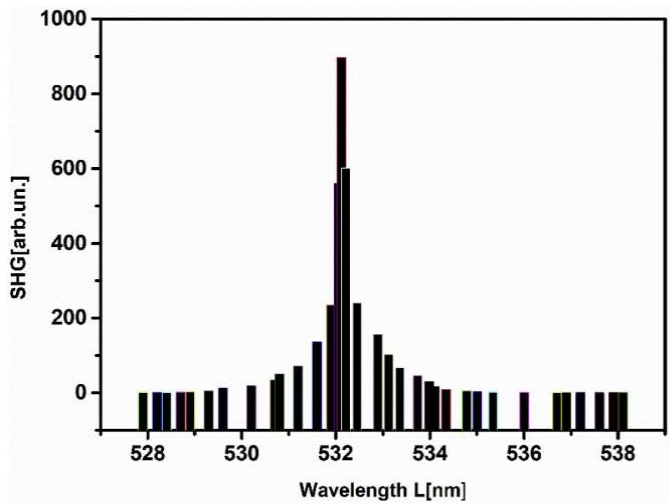

**Fig. 8.** The spectrum of the SHG after two beams of coherent treatment.

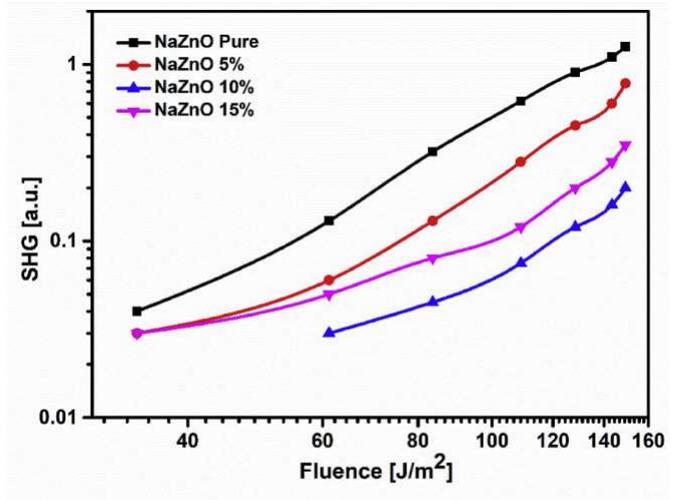

(a)

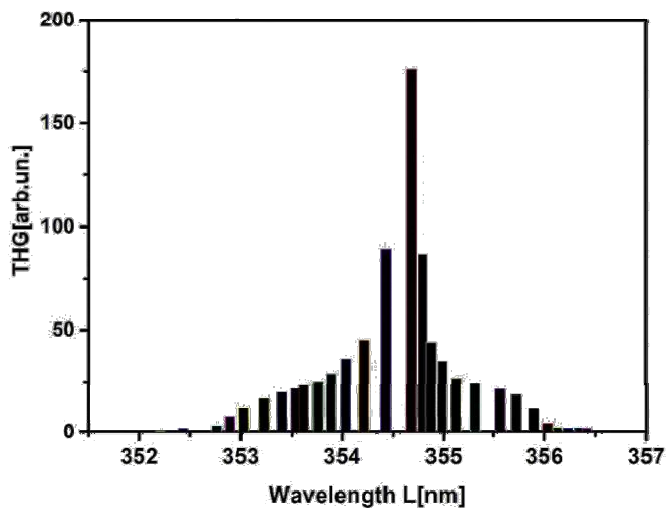

**Fig. 9.** The spectrum of the THG after two-beam coherent treatment.

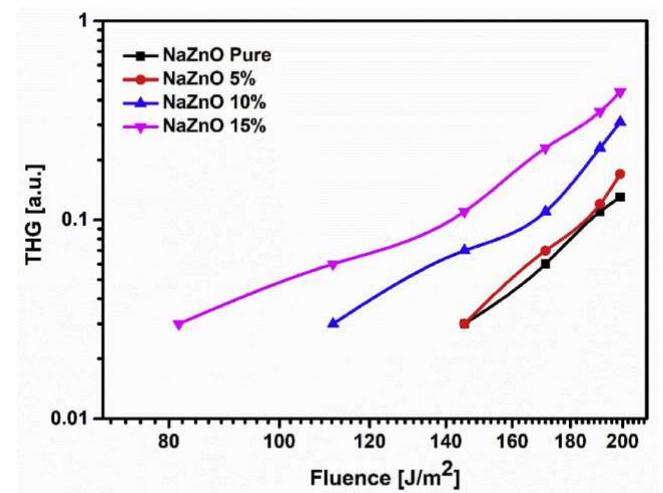

(b)

**Fig. 10.** (a) SHG and (b) THG stimulated by the fundamental laser light.

Since the excitation source used in the experiment were a continuous mode operated laser source the nonlinear absorption process exhibited by Na: ZnO can be attributed to a thermal nonlinear effect. The scattered experimental points observed were due to the presence of nonlinear scattering mechanism exhibited by the films at the waist region. The nonlinear absorption due to two photon absorption mechanism which is predominant with slight nonlinear scattering. Hence, at the waist region where the intensity is high, the variations in experimental and theoretical nonlinear absorption spectra arises due to nonlinear scattering effect. Table 4 shows the variation in the $\beta_{eff}$ with respect to Na doping concentration. Table 5 shows the comparison of nonlinear absorption coefficient values of nearly reported Semiconductor thin films.

*3.4.2. Photostimulated second and third harmonic generation measurements*

The laser stimulated nonlinear optical effects which is principally different with respect to the classical nonlinear optical phenomena described in sec 3.4.1 were explored. These effects are very important for designing laser operated devices like optical triggers, modulators and reflectors. The principal set-up for the measurements of the coherent laser stimulated second and third harmonic generation is presented in Fig. 7.

A 10 ns Nd:YAG fundamental pulsed laser emitting at 1064 nm with a pulse duration of 8–10 ns and pulse frequency repetition 10 Hz were used as a basic source of the laser stimulated nonlinear optical studies. The Glan polarizer was used to vary the input beam energy density of the fundamental beam. The semi-transparent mirrors have been used for the space splitting and space separation of the fundamental beam for two coherent channel. One of the channels has been served (at the same frequency) for the monitoring of the fundamental laser beam control and another one for the probing of the second and third harmonic generation after their spectral separation by interferometer filters at 532 nm and 355 nm respectively. Additionally, the spectral shape of the doubled and tripled frequency beam has been detected by spectrometer DFS8 (with diffraction gratings) with spectral resolution about 0.2 nm. The 1ns relaxed photomultiplier was put after the interferometric filters at 532 nm and 355 nm and output slit of the spectrometer with a band path about 5 nm. This was used for the spectral separation of the SHG and THG signal with respect to the fundamental light and fluorescence parasitic background. The incident fundmental laser beam has been space separated which allow to perform the photoinduced treatment during several seconds to form some anisotropic grating similarly to the described in ref. [30] So we have coherent interactions between the laser stimulated beam effects and the output NLO.

The typical spectra of the SHG and THG are shown in Figs. 8 and 9. One can clearly see a drastic jump in the output of SHG and THG with respect to the background which unambiguously confirms the nonlinear optical origin of the observed effects. The data are obtained after the initial treatment by 50–70 two coherent pulses separated in the space and incident on the sample surface at 25–30°. The maximal signal was observed for 45-degree angles between the two photo induced beams. The SHG output signals after treatments have been increased up to 22–25% and of THG about 10%. The BiB$_3$O$_6$ pure and doped crystallite powders [28] and the oxide glasses have been used for the reference to evaluate the efficiency of the output SHG and ZnSe:Co$^{2+}$ crystals for the THG signals [29]. These experiments have been performed in more than 30 different points of the specimen's to avoid the non-homogeneity of the samples as two different stage. The photo treatment has caused a saturation of the output SHG and THG and was controlled by the stabilization of the beam grating pattern profiles using the CCD camera. The principal results of the treatment versus the fundamental power density are given in Fig. 10. Following Fig. 10 one can clearly see that the doping by Na leads to decrease of the SHG efficiency with respect to the pure ZnO samples.

The presented results clearly show that the maximal signal of the SHG is observed for the pure ZnO thin films. The maximal suppression is observed for the 10% Na doping. At the same time at 15% just higher. It may reflect an occurrence of partial aggregation of the Na and fur-thermore, at 5% the signal intensity is higher than for the 10% and 15% Na doping. It is necessary to empathize that because the effects are laser stimulated there is lacking a possibility to analyze the power depen-dence of the samples for the pure nonlinear optical effects.

The SHG/THG presented in Fig. 9 demonstrate different non-monotonous dependence of signal intensity with Na doping concentration. For pure ZnO, the THG signal is minimal and the maximum is observed for the 15% Na doping. So it is quite different with respect to the SHG behaviour. This may be explained by different mechanisms of the SHG and THG. For the SHG principal role belongs to the charge density eccentricity and for the THG principal role is contributed by the changes in the ground state dipole moments.

To understand the laser stimulated dependences of the THG and SHG it is necessary to take into account different phenomenon with respect to the traditional NLO effects. In this case the general phenomenology of the NLO effect will be different and may be described by a phenomenology modifed by exernal laser induced field:

$$P_i^{(2)}(\omega) = d_{ijk}^{(2)}(2\omega; \omega, \omega, 0) E_j(\omega) E_k(\omega) E_{ef}(0) \quad (4)$$

$$P_i^{(3)}(\omega) = d_{ijkl}^{(3)}(3\omega; \omega, \omega, \omega, o) E_j(\omega) E_k(\omega) E_l(\omega) E_{ef}(0), \quad (5)$$

where $d_{ijk}$ correspond to the nonlinear optical susceptiblies of the second and third order described by third and fourth order polar tensors for SHG and THG, respectively. Principal difference with respect to traditional NLO phenomenological description consists in the occur-rence of the internal dc-electric field $E_{ef}$ which is due to interaction of the two coherent beams as described in ref. [30]. Additionally due to photoinduced changes some modification will be appeared in the co-herent lengths due to laser induced birefringence. As a consequence the power dependences of the SHG/THG will be higher than 2 and 3 as for the traditional SHG/THG.

Additionally, the space profile of the beams has been controlled using the CCD and the results are presented in Figs. 11–13. One can see that initial beam profile was disturbed by grating interference patterns which means that the effects of coherent gratings appear. One can see that for both cases there exist two types of grating patterns in the profiles for the beams. The frequencies of the SHG patterns are less than for the THG. It may reflect different mechanisms of the matching for these two cases.

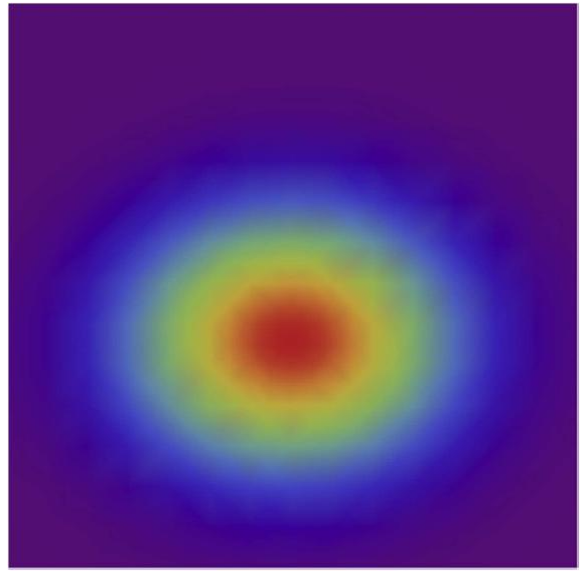

**Fig. 11.** Space profile of the incident beam computer reconstructed from CCD camera.

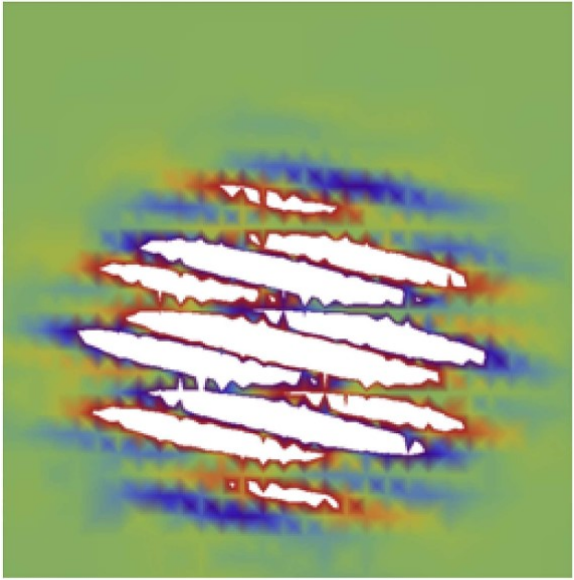

**Fig. 12.** Space profile of the output SHG reconstructed using CCD.

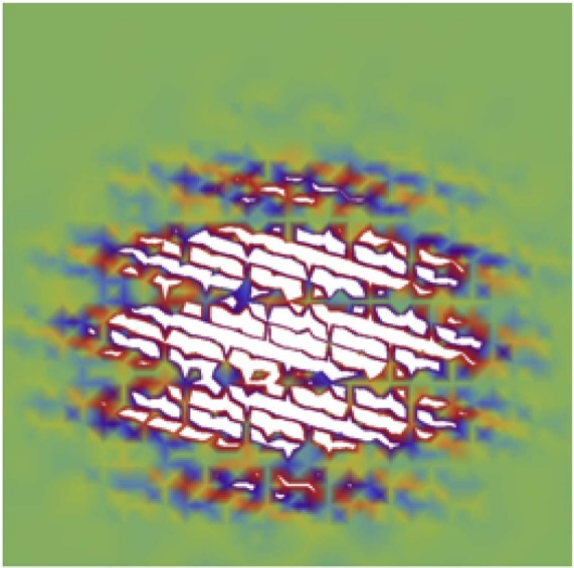

**Fig. 13.** Space profile of the output THG reconstructed using CCD.

### 4. Conclusions

Na:ZnO thin film at different concentration have been prepared by spray pyrolysis technique. The structural, optical and nonlinear optical properties of the thin film were studied. XRD analysis reveals grown films exhibits a c-axis orientation growth along (002) plane. The PL studies show three emission centres at violet, blue, green region due to the presence of native defect sites. Open aperture Z-scan curve indicates a switching over behaviour from RSA to SA and vice versa upon Na incorporation. Maximal signal of the SHG was observed for the pure ZnO nanofilms whereas 15% Na:ZnO thin films exhibited maximum THG signal. The maximal suppression of SHG was observed for the 10% Na doping. It is necessary to emphasize that during laser stimulated effects we deal with principally different phenomenon with respect to the traditional NLO effects. In this case the general phenomenology of the NLO effect will be different and may be described by a phenom-enology modifed by exernal laser induced field. The power of NLO coefficient dependences will be different form 2 and 3 in the case of the SHG and THG, respectively. This fact may be used for the formation of the multi-functional nonlinear optical devices and for the tuning of nonlinear optical susceptibilities by appropriate changes in the laser-stimulated beams and doping concentrations.

### Acknowledgements

Presented results are part of a project that has received funding from the European Union's Horizon 2020 research and innovation programme under the Marie Skłodowska-Curie grant agreement No 778156.Supported from resources for science in years 20182022 granted for the realization of international co-financed project Nr W13/H2020/2018 (Dec. MNiSW 3871/H2020/2018/2) is also acknowledged."

One of the author Manjunatha K.B. would like to thank Vision Group on Science &Technology, Department of I. T, B.T and S&T, Government of Karnataka for supporting this work through the SMYSR scheme.

### References

[1] D. Akcan, A. Gungor, L. Arda, Structural and optical properties of Na-doped ZnO Films, J. Mol. Struct. 1161 (2018) 299–305.
[2] O.W. Kennedy, M.L. Coke, E.R. White, M.S.P. Shaffer, P.A. Warburton, MBE growth and morphology control of ZnO nanobelts with polar axis perpendicular to growth direction, Mater. Lett. 212 (2018) 51–53.
[3] R. Mahdavi, S.S. Ashraf Talesh, The effect of ultrasonic irradiation on the structure, morphology and photocatalytic performance of ZnO nanoparticles by sol-gel method, Ultrason. Sonochem. 39 (2017) 504–510.
[4] N. Neamjan, W. Sricharussin, P. Threepopnatkul, Effect of various shapes of ZnO nanoparticles on cotton fabric via electrospinning for UV-blocking, J. Nanosci. Nanotechnol. 12 (2012) 525–530.
[5] A. Douayar, M. Abd-Lefdil, K. Nouneh, P. Prieto, R. Diaz, A.O. Fedorchuk, I.V. Kityk, PhotoinducedPockels effect in the Nd-doped ZnO oriented nanofilms, Appl. Phys. B 110 (3) (2013) 419–423.
[6] N.S. AlZayed, JeanMichel JeanEbothé, I.V. Kityk, O.M. Yanchuk, D.I. Prots, O.V. Marchuk, Influence of the ZnO nanoparticles sizes and morphology on the photoinduced light reflectivity, Physica E60 (2014) 220–223.
[7] Mohamed a. Basyooni, Mohamed Shaban, Adel M. El Sayed, Enhanced gas sensing properties of spin-coated Na-doped ZnO nanostructured films, Sci. Rep. 7 (2017) 41716.
[8] L.W. Wang, F. Wu, D.X. Tian, W.J. Lia, L. Fang, C.Y. Kong, M. Zhou, Effect of Na content on structural and optical properties of Na-doped ZnO thin films prepared by sol-gel method, J. Alloy. Comp. 623 (2015) 367–373.
[9] T. Srinivasulu, K. Saritha, K.T. Ramakrishna Reddy, Synthesis and characterization of Fe-doped ZnO thin films deposited by chemical spray pyrolysis, Modern Electr. Mater. 3 (2017) 76–85.
[10] I.V. Kityk, J. Ebothe, A. Elhichou, Pressure-temperature anomalies of doped ZnO polycrystalline films deposited on bare glasses, Mater. Lett. 51 (6) (2001) 519–524.
[11] I.V. Kityk, A. Migalska-Zalas, J. Ebothe, A. Elchichou, M. Addou, A. Bougrine, A.Ka. Chouane, Anomalously large pockels effect in ZnO-F single crystalline films deposited on bare glass, Cryst. Res. Technol. 37 (4) (2002) 340–352.
[12] J. Ebothe, W. Gruhn, A. Elhichou, I.V. Kityk, R. Dounia, M. Addou, Giant piezo optics effect in the ZnO–Er3+ crystalline films deposited on the glasses, Optic Laser. Technol. 36 (2004) 173–180.
[13] J. Ebothe, R. Miedzinski, V. Kapustianyk, B. Turko, B. Kulyk, W. Gruhn, I.V. Kityk, Optical SHG for ZnO films with different morphology stimulated by UV-laser treatment, J. Phys. Conf. Ser. 79 (2007) 012001.
[14] A. Douayar, M. Abd-Lefdil, K. Nouneh, P. Prieto, R. Diaz, A.O. Fedorchuk, I.V. Kityk, PhotoinducedPockels effect in the Nd-doped ZnO oriented nanofilms, Appl. Phys. B 110 (3) (2013) 419–423.
[15] Lawrence K. Dintle, Pearson V.C. Luhanga, Charles Moditswe, Cosmas M. Muiva, Compositional dependence of optical and electrical properties of indium doped zinc oxide(IZO) thin films deposited by chemical spray pyrolysis, Low-dimens. Syst. Nanostruct. 99 (2018) 91–97.
[16] Okba Belahssen, Hachemi Ben Temam, Said Lakel, Boubaker Benhaoua, Said Benramache, salim gareh, Effect of optical gap energy on the Urbach energy in the undoped ZnO thin films, Optik 126 (2015) 1487–1490.
[17] Saliha ILICAN, Effect of Na doping on the microstructures and optical properties of ZnO nanorods, J. Alloy. Comp. 553 (2013) 225–232.
[18] Changle Wu, Qingli Huang, Synthesis of Na-doped ZnO nanowires and their pho-tocatalytic properties, J. Lumin. 130 (2010) 2136–2141.
[19] Albin Antony, P. Poornesh, K. Ozga, J. Jedryka, P. Rakus, I.V. Kityk, Enhancement of the efficiency of the third harmonic generation process in ZnO:F thin films probed by photoluminescence and Raman spectroscopy, Mater. Sci. Semicond. Process. 87 (2018) 100–109.
[20] M. Sheikh-bahae, A.A. Said, T.-H. Wei, D.J. Hagan, E.W.V. Stryland, Sensitive measurement of optical nonlinearities using a single beam, IEEE J. Quantum Electron. 26 (1990) 760–769.

[21] Albin Antony, P. Poornesh, I.V. Kityk, G. Myronchuk, Ganesh Sanjeev, Vikash Chandra Petwal, Vijay Pal Verma, Jishnu Dwivedi, A study of 8MeVe-beam on localized defect states in ZnO nanostructures and its role on Photoluminescence and third harmonic generation, J. Lumin. 207 (2019) 321–332.

[22] M.S. Bannur, Albin Antony, K.I. Maddani, P. Poornesh, Ashok Rao, K.S. Choudhan "Tailoring the Nolinear Optical Susceptibility, Photoluminescence and Optical Band Gap of Nanostructured SnO2 Thin Fims by Zn Doping for Photonic Device.

[23] L. Castaneda, O.G. Morales-Saavedra, D.R. Acosta, A. Maldonado, Structural, morphological, optical, and nonlinear optical properties of fluorine-doped zinc oxide thin films deposited on glass substrates by the chemical spray technique, Phys.Stat. Sol.(a) 203 (2006) 1971.

[24] Y.S. Tamgadge, A.L. Sunatkari, S.S. Talwatkar, V.G. Pahurkar, G.G. Muley, Linear and nonlinear optical properties of nanostructured Zn(1x)SrxO–PVA composite thin films, Opt. Mater. 37 (2014) 42–50.

[25] Z. Dehghani, S. Nazerdeylami, E. Saievar-Iranizad, M.M. Ara, Synthesis and investigation of nonlinear optical properties of semiconductor ZnS nanoparticles, J. Phys. Chem. Solids 72 (2011) 1008–1010.

[26] S. Nazerdeylami, E. Saievar-Iranizad, Z. Dehghani, M. Molaei, Synthesis and pho-toluminescent and nonlinear optical properties of manganese doped ZnS nano-particles, Phys. B Condens. Matter 406 (2011) 108–111.

[27] V. Ganesh, I.S. Yahiaa, S. AlFaify, Mohd. Shkir, Sn-doped ZnO nanocrystalline thin films with enhanced linear and nonlinear optical properties for optoelectronic ap-plications, J. Phys. Chem. Solids 100 (2017) 115–125.

[28] A. Brenier, I.V. Kityk, A. Majchrowski, Evaluation of $Nd^{3+}$-doped $BiB_3O_6$ (BIBO) closed with Pr3+ doped BiB3O6 as a new potential self-frequency conversion laser crystal, Optic Commun. V. 203 (2002) 125–132.

[29] Z. Mierczyk, A. Majchrowski, K. Ozga, A. Slezak, I.V. Kityk, Simulation of nonlinear optical absorption in ZnSe:Co2+ crystals, Optic Laser. Technol. 38 (7) (2006) 558–564.

[30] M.K. Balakirev, I.V. Kityk, V.A. Smirnov, L.I. Vostrikova, J. Ebothe, Anisotropy of the optical poling of glass, Phys. Rev. A67 (2003) 023806.